\documentclass[twocolumn,showpacs,preprintnumbers,superscriptaddress,amsmath,amssymb]{revtex4}

\usepackage{graphicx}
\usepackage{dcolumn}
\usepackage{bm}
\begin{document}

\preprint{}

\title{Spectral Measure of Robustness in Complex Networks} with \\

\author{Jun Wu}
\email{wujunpla@hotmail.com}
\affiliation{College of Information Systems and Management, National University of Defense Technology, Changsha 410073, P. R. China.}
\affiliation{Institute of Computing Technology, Chinese Academy of Sciences, Beijing 100080, P. R. China}
\affiliation{Institute for Mathematical Sciences, Imperial College London, London SW7 2PG, United Kingdom}
\author{Yue-Jin Tan}
\author{Hong-Zhong Deng}
\author{Yong Li}
\author{Bin Liu}
\author{Xin Lv}

\affiliation{College of Information Systems and Management, National University of Defense Technology, Changsha 410073, P. R. China.}

\date{\today}

\begin{abstract}
We introduce the concept of natural connectivity as a robustness measure of complex networks. The natural connectivity has a clear physical meaning and a simple mathematical formulation. It characterizes the redundancy of alternative paths by quantifying the weighted number of closed walks of all lengths. We show that the natural connectivity can be derived mathematically from the graph spectrum as an average eigenvalue and that it increases strictly monotonically with the addition of edges. We test the natural connectivity and compare it with other robustness measures within a scenario of edge elimination. We demonstrate that the natural connectivity has an acute discrimination which agrees with our intuition.
\end{abstract}


\maketitle We are surrounded by networks. Networks with complex topology describe a wide range of systems in nature and society. The study of complex networks has become an important area of multidisciplinary research involving physics, mathematics, biology, social sciences, informatics, and other theoretical and applied sciences \cite{Newman, Boccaletti}. Complex networks rely for their function and performance on their robustness, i.e., the ability to endure threats and survive accidental events. For example, modern society is dependent on its critical infrastructure networks: communication, electrical power, rail, and fuel distribution networks. Failure of any of these critical infrastructure networks can bring the ordinary activities of work and recreation to a standstill. Other examples of robustness arise in biological and social systems, including questions such as the stability of social organizations in the face of famine, war, or even changes in social policy. Because of its broad application, robustness has become a central topic in all complex networks and receives growing attention.

As a basic concept of graph theory, the connectivity of a graph is an important and probably the earliest measure of robustness of a network \cite{Frank}. Vertex (edge) connectivity, defined as the size of the smallest vertex (edge) cut, determines in a certain sense the robustness of a graph to the deletion of vertices (edges). However, the vertex or edge connectivity only partly reflects the ability of graphs to retain certain degrees of connectedness after deletion. Other improved measures were introduced and studied, including super connectivity \cite{Bauer}, conditional connectivity \cite{Harary}, restricted connectivity \cite{Esfahanian}, fault diameter \cite{Krishnamoorthy}, toughness \cite{Chvatal}, scattering number \cite{Jung}, tenacity \cite{Cozzen}, expansion parameter
\cite{Alon} and isoperimetric number \cite{Mohar}. In contrast to vertex (edge) connectivity, these new measures consider both the cost to damage a network and how badly the network is damaged. However, from an
algorithmic point of view, it is unfortunate that the problem of calculating
these measures for general graphs is NP-complete. This implies that these measures are of no great use within the context of complex networks.

Another remarkable measure to unfold the robustness of a network is the second smallest eigenvalue of the Laplacian matrix, also known as the algebraic connectivity. Fiedler \cite{Fiedler} showed that the magnitude of the algebraic connectivity reflects how well connected the overall graph is, i.e., the larger the algebraic connectivity is, the more difficult it is to cut a graph into independent components. Hence, there is a vast literature on the algebraic connectivity (see \cite{Merris} for a survey). However, the algebraic connectivity
is equal to zero for all disconnected networks. Therefore, it is too coarse
a measure for complex networks.

The prime study regarding network robustness within the context of complex networks came from random graph theory \cite{Bollobas} and was stimulated by the work of Albert et al. \cite{Albert}. Instead of a strict extremal property, they proposed a statistical measure, i.e., the critical removal fraction of vertices (edges) for the disintegration of a network, to characterize
the robustness of complex networks. The disintegration of networks can be observed from the decrease of network performance. The most common performance measurements include the diameter, the size of largest component, the average path length, the efficiency \cite{Latora,Crucitti} and the number of reachable vertex pairs \cite{Palmer,Siganos}. As the fraction of removed vertices or edges increases, the network will eventually collapse at a critical fraction. It is suggested that scale-free networks display an exceptional robustness against random failure, but show poor performance against intentional attack \cite{Albert}. As an expansion of the work by Albert et al., Wu et al. \cite{wu1, wu2} studied the robustness of complex networks under incomplete information, i.e., one can only obtain the information of partial vertices. Cohen et al.
\cite{Cohen1, Cohen2} developed the first analytical approach to calculating the critical removal fraction of a network under random failure or intentional attack. Callaway et al. \cite{Callaway} put forward an alternative and more general approach using a generalization of the generating function formalism.

If we consider a source vertex and a termination vertex, there may be several alternative paths between them. When one path fails, the two vertices can still communicate through other alternative paths. It is intuitive that the more alternative paths, the more robust the connectedness between the two
vertices. This observation leads us to consider the redundancy of alternative paths as the root of the robustness of networks, which ensures that the connection between vertices still remains possible in spite of damage to the network.
Although it would be ideal to define this redundancy as the number of alternative paths of different lengths for all pairs of vertices, this measure is very difficult to calculate. Note, however, that the number of closed walks in a network is a good index for the number of alternative paths. In this paper, we propose a new robustness measure of complex networks based on the number of closed walks. 

A complex network can be viewed as a simple undirected graph $G(V,E)$, where $V$ is the set of vertices, and $E \subseteq V \times V$ is the set of edges. Let $N = \left| V \right|$ and $M = \left| E \right|$ be the number of vertices and the number of edges, respectively. Let $d_i $ be the degree of vertex $v_i$. Let $d_{\min } $ be the minimum degree and $d_{\max } $ be the maximum degree of $G$. Let $A(G) = (a_{ij} )_{N \times N} $ be the adjacency matrix of $G$, where $a_{ij}  = a_{ji}  = 1$ if vertex $v_i $ and $v_j $ are adjacent, and $a_{ij}  = a_{ji}  = 0$ otherwise. A walk of length $k$ in a graph $G$ is an alternating sequence of vertices and edges $v_0e_1v_1e_2...e_kv_k$, where $v_i \in V$ and $e_i=(v_{i-1}, v_i) \in E$. A walk is closed if $v_0=v_k$.

Closed walks are directly related to the subgraphs of the graph. For instance, a closed walk of length $k=2$ corresponds to an edge and a closed walk of length $k=3$ represents a triangle. Note that a closed walk can be trivial, i.e., containing repeated vertices, leading to the length of a closed walk being infinite. The number of closed walks is an important index for complex networks. For example, Estrada et al. have measured vertex centrality \cite{Estrada1} and network bipartivity \cite{Estrada2} based on the number of closed walks. Here we define the redundancy of alternative paths as the number of closed walks of all lengths. Considering that shorter closed walks have more influence on the redundancy of alternative paths than longer closed walks and to avoid the number of closed walks of all lengths to diverge, we scale the contribution of closed walks to the redundancy of alternative paths by dividing them by the factorial of the length $k$. That is, we define a weighted sum of numbers of closed walks $S =\sum\nolimits_{k = 0}^\infty {\left( {n_k /k!} \right)} $, where $n_k $ is the number of closed walks of length $k$. Using matrix theory, we know that
 \begin{equation}
\label{nk}  n_k  = \sum\limits_{i_1 ,i_2 ,...i_k } {a_{i_1 i_2 } }
a_{i_2 i_3 } ...a_{i_k i_1 }  = trace(A^k ) = \sum\limits_{i = 1}^N {\lambda _i^k},
\end{equation} where $\lambda _i $ is the $i$th largest eigenvalue of $A(G)$. Specifically, $n_2  = \sum\nolimits_i {d_i }  = 2M$. Using Eq.~(\ref{nk}),
we obtain
\begin{equation}
\label{S}
S = \sum\limits_{k = 0}^\infty  {\frac{{n_k }}{{k!}}}  =
\sum\limits_{k = 0}^\infty  {\sum\limits_{i = 1}^N {\frac{{\lambda
_i ^k }}{{k!}}} }  = \sum\limits_{i = 1}^N {\sum\limits_{k =
0}^\infty  {\frac{{\lambda _i ^k }}{{k!}}} }  = \sum\limits_{i =
1}^N {e^{\lambda _i } } .
\end{equation}

Eq.~(\ref{S}) shows that the weighted sum of closed walks of all lengths can be derived from the graph spectrum. Noting that $S$ will be a large number for large $N$, we scale $S$ and denote it by $\bar \lambda $
\begin{equation}
\label{S2}
 \bar \lambda = \ln \left( {\frac{{S }}{N}} \right)=\ln
\left( {\frac{{\sum\limits_{i = 1}^N {e^{\lambda _i } } }}{N}}
\right),
\end{equation}
which corresponds to an 'average eigenvalue'. We propose to call it \emph{natural connectivity} or \emph{natural eigenvalue}.

A desired property of natural connectivity is that it changes monotonically when edges are added or deleted. Let $G + e $ be the graph obtained by adding an edge $e$ to $G$ and let  $\hat n_k  = \hat n'_k  + \hat n''_k $ be the number of closed walks of length $k$ in $G + e $, where $ \hat n'_k $ is the number of closed walks of length $k$ containing $e$ and $ \hat n''_k $ is the number of closed walks of length $k$ containing no $e$. Note that $\hat n'_k = n_k$ and $ \hat n''_k \geq 0$, thus $\hat n_k \ge n_k$. It is easy to show that $\hat n_k > n_k$ for some $k$, e.g., $\hat n_2= n_2+2$. Consequently, $\bar \lambda (G + e) > \bar \lambda (G) $, indicating that the natural connectivity increases strictly monotonically as edges are added. In Fig. 1, we illustrate two simple graphs with six vertices, where graph (b) is obtained by adding an edge to graph (a). Our intuition suggests that graph (b) should be more robust than graph (a), which agrees with our measure. The natural connectivity of graph (a) and (b) are 1.0878 and 1.3508, respectively. However, some robustness measures mentioned above can not distinguish the two graphs. For example, both graphs have identical edge connectivity 2 and identical algebraic connectivity 0.7369.

\begin{figure}[htbp]
 \centering
  \includegraphics[width=0.25\textwidth]{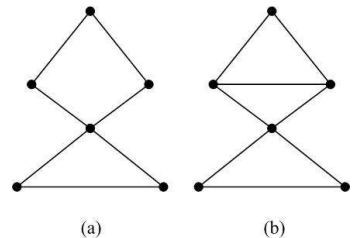}
\caption{Graph (b) is obtained from graph (a) by adding an edge. Both graphs have identical edge connectivity and identical algebraic connectivity, but are distinguished by our proposed natural connectivity.}
\label{fig1}
\end{figure}

It is evident from Eq.~(\ref{S2}) that $\lambda _1  \ge \bar
\lambda \ge \lambda _N $. Moreover, for a given number of vertices $N$ and
following the discussion on monotonicity above, the empty graph consisting of isolated vertices has the minimum natural connectivity and the complete graph, whose vertices are pairwise adjacent, has the maximum natural connectivity. It is known that $\lambda _1  = \lambda _2  = ... = \lambda _N  = 0$ for the empty graph, and $\lambda _1  = N - 1, \lambda _2  = \lambda _3  = ... = \lambda _{N }  =  - 1 $  for the complete graph \cite{Cvetkovic}. Hence we obtain the following bound for the natural connectivity
\begin{equation}
\label{bound} 0 \le \bar \lambda  \le \ln ((N - 1)e^{ - 1}  + e^{N
- 1} ) - \ln N \approx N-\ln N.
\end{equation}

To explore in depth the natural connectivity measure and compare it with
other robustness measures, we consider a scenario of edge elimination. As
edges are deleted, we expect the decrease of the robustness measure, and we also expect different behavior for different edge elimination strategies. We generate initial networks with a power-law degree distribution using the BA model \cite{Barabasi}. We remark that the type of network has no effect
on the analysis and conclusions. We consider four edge elimination strategies:
(i) deleting the edges randomly (random strategy); (ii) deleting the edges connecting high-degree vertices and high-degree vertices in the descending order of $d_i \cdot d_j $, where $d_i $ and $d_j $ are the degrees of the end vertices of an edge (rich-rich strategy); (iii) deleting the edges connecting low-degree vertices and low-degree vertices in the ascending order of $d_i \cdot d_j $ (poor-poor strategy); (iv) deleting the edges connecting high-degree vertices and low-degree vertices in the descending order of $\left| {d_i  - d_j } \right|$ (rich-poor strategy). Along with the natural connectivity, we investigate  three other robustness measures: edge connectivity $ \kappa _{\rm{E}} (G) $, algebraic connectivity $a(G)$ and critical removal fraction of vertices under random failure $f_c^R $. To find the critical removal fraction of vertices, we choose $\kappa \equiv <k^2 >/<k>\leq 2$ as the criterion for the disintegration of networks \cite{Cohen1}. The results are shown in Fig. 2. Each measure is an average over 100 realizations of a BA network.

\begin{figure}[htbp]
 \centering
  \includegraphics[width=0.4\textwidth]{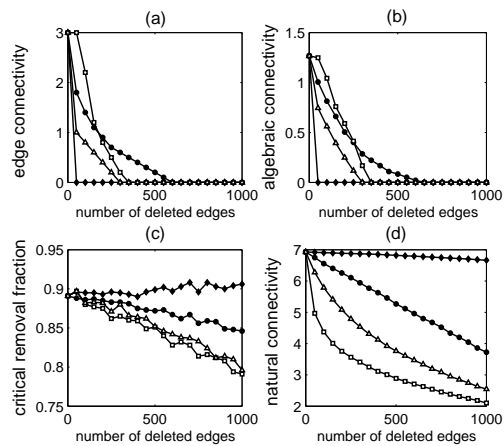}
\caption{The robustness measured by edge connectivity (a), algebraic connectivity
(b), critical removal fraction of vertices (c) and natural connectivity (d) as a function of number of deleted edges for four edge elimination strategies: random strategy (circles), rich-rich strategy (squares), poor-poor strategy
(diamonds) and rich-poor strategy (triangles). The initial network is generated using the BA model, where $N=1000$ and $<k>\approx6$. Each quantity is an average over 100 realizations.} \label{fig2}
\end{figure}

From Fig. 2(a) and Fig. 2(b), we find similar behaviour between $ \kappa _{\rm{E}} (G) $ and $a(G)$. The first observation is that deleting a small
quantity of edges connecting high-degree vertices and high-degree vertices has no obvious effect on the robustness measured by the edge or algebraic connectivity. On the other hands, the robustness drops rapidly under the poor-poor strategy, i.e., when small quantity of edges connecting low-degree vertices and low-degree vertices are deleted. It is generally believed that the edges between high-degree vertices are important, and the edges between low-degree vertices are inessential for the global network robustness. For example, in the Internet, the failure of the links between core routers will bring a disaster, but there is no effect on the network robustness if we disconnect two terminal computers. Clearly, robustness measures
based on edge or algebraic connectivity do not agree with our intuition. These unexpected features can be explained by the bound $a(G)\le \kappa(G) \le \kappa _{\rm{E}} (G) \le d_{\min } $, also known as Fiedler's inequality \cite{Fiedler}, where $\kappa(G)$ is the vertex connectivity. In fact, we find that the probability of $ \kappa {}_{\rm{E}}(G) =d_{\min } $ almost approaches to 1. After a few edges connecting low-degree vertices and low-degree vertices are deleted, $d_{\min } $ decreases to zero rapidly, but $ d_{\min } $ is preserved under rich-rich strategy. Moreover, we find that, for all four strategies, the edge or algebraic connectivity is equal to zero after particular edges are deleted, even in the case where only very few vertices are separated from the largest cluster. This means that both the edge connectivity and the algebraic connectivity lose discrimination when the network is disconnected.

Figure 2(c) shows the critical removal fraction of vertices $f_c^R $ as a function of the number of deleted edges. Contrary to the result of edge or algebraic connectivity and in agreement with our intuition, we observe that the rich-rich strategy is the most effective edge elimination strategy and the poor-poor strategy is the worst one. Nevertheless, we find that there are irregular fluctuations in the curves. This shows that the critical removal fraction is not a sensitive measure of robustness, especially for small sized
networks.

In Fig. 2(d), we display the results of the natural connectivity according
to Eq.~(\ref{S2}). We find a clear variation of the measure with distinct differences between the four edge elimination strategies, showing a clear ranking for the four edge elimination strategies: rich-rich strategy $\succ$ rich-poor strategy $\succ$ random strategy $\succ$ poor-poor strategy, which agrees with our intuition. For the random strategy, we observe a linear decrease of the natural connectivity. For the rich-rich strategy or rich-poor strategy, the natural connectivity decreases rapidly with the edge elimination. For poor-poor strategy, deleting a small quantity of edges connecting low-degree vertices and low-degree has weak effect on the robustness. Moreover, we find that the curves for natural connectivity are surprisingly smooth, which indicates that the natural connectivity can measure the robustness of complex networks stably even for very small sized networks. In fact, we have found that the curves for natural connectivity are also smooth  without averaging over 100 realizations, viz. for one individual network. However, in the case of individual networks, we find stepped curves for the edge or algebraic connectivity and
large fluctuations for the critical removal fraction.

In summary, we have proposed the concept of natural connectivity as a spectral measure of robustness in complex networks. The natural connectivity is rooted in the inherent structural properties of a network. The theoretical motivation of our measure arises from the fact that the robustness of a network comes from the redundancy of alternative paths. The natural connectivity is expressed in mathematical form as an average eigenvalue and allows a precise quantitative analysis of the network robustness. Our measure works both in connected and disconnected networks. We have shown that it changes strictly monotonically with the addition or deletion of edges. To test our natural connectivity measure and compare it with other measures, we have designed a scenario of edge elimination, in which four different edge elimination strategies are considered. We have demonstrated that the natural connectivity has an acute discrimination in measuring the robustness of complex networks and can detect
small variations of robustness stably. Rich information about the topology and dynamical processes can be extracted from the spectral analysis of the networks. The natural connectivity sets up a bridge between graph spectra and the robustness of complex networks. It is of great theoretical and practical significance in network design and optimization to link the robustness to other network structural or dynamical properties (e.g., efficiency, synchronization, diffusion, searchability).

\begin{acknowledgments}
We would like to thank Mauricio Barahona and Quoqing Zhang for useful
discussions and comments. This work is in part supported by the National
Science Foundation of China under Grant No. 70501032, No. 70771111
and No. 60673168. This work is also partly supported by the
Hi-Tech Research and Development Program of China under Grant No.
2006AA01Z207.
\end{acknowledgments}

\end{document}